\documentclass[aps,prl,twocolumn,groupedaddress,amsmath,amssymb]{revtex4-1}
\usepackage{graphicx}  
\usepackage{dcolumn}   
\usepackage{bm}        
\usepackage{verbatim}   
\usepackage{hyperref}
\usepackage[usenames,dvipsnames]{color}
\hypersetup{colorlinks,citecolor=Blue,linkcolor=Red,urlcolor=Blue}
\usepackage[titletoc]{appendix}
\usepackage{amsmath}
\usepackage{natbib}

\begin{document}

\title{Resonant Activation of Population Extinctions}
\author{Christopher Spalding}
\affiliation{
   Divison of Geological and Planetary Sciences,\\
   California Institute of Technology,\\
   Pasadena, CA 91125
   }
\author{Charles R. Doering}
\affiliation{Center for the Study of Complex Systems, University of Michigan, Ann Arbor, Michigan 48109-1107\\
Department of Mathematics, University of Michigan, Ann Arbor, Michigan 48109-1043 and\\
 Department of Physics, University of Michigan, Ann Arbor, Michigan 48109-1040}
   
   \author{Glenn R. Flierl}
\affiliation{Department of Earth, Atmospheric, and Planetary Sciences, Massachusetts Institute of Technology, Cambridge, Massachusetts 02139-4307}
\date{\today}
\begin{abstract}
Understanding the mechanisms governing population extinctions is of key importance to many problems in ecology and evolution. Stochastic factors are known to play a central role in extinction, but the interactions between a population's demographic stochasticity and environmental noise remain poorly understood. Here, we model environmental forcing as a stochastic fluctuation between two states, one with a higher death rate than the other. We find that in general there exists a rate of fluctuations that minimizes the mean time to extinction, a phenomenon previously dubbed ``resonant activation." We develop a heuristic description of the phenomenon, together with a criterion for the existence of resonant activation. Specifically the minimum extinction time arises as a result of the system approaching a scenario wherein the severity of rare events is balanced by the time interval between them. We discuss our findings within the context of more general forms of environmental noise, and suggest potential applications to evolutionary models.
 \end{abstract}

\maketitle
 \section{Introduction}
The extinction of populations, and of entire species, has played a critical role in shaping global biodiversity \citep{Macarthur1967,Bambach2006}. In more recent times, human impacts upon extinction risks are being felt at an increased rate \citep{Ceballos2015}. Accordingly, in order to better understand the history of life and to reliably forecast future ecological crises, it is crucial to understand the mechanisms governing extinction. 

Some of the earliest attempts to quantitatively model biological populations treated the number of individuals as a continuous variable, evolving deterministically as the difference between prescribed birth rates and death rates \citep{Lotka1920}. The ``carrying capacity" $K$ was typically defined as the population size at which births and deaths are equal. Deterministic approaches have proved beneficial in understanding numerous qualitative features of population sizes. However, in order to predict the time over which a population is likely to go extinct, a stochastic treatment is required \citep{Kendall1949,Doering2005,Melbourne2008}. 

The mean extinction times of populations under the action of various forms of stochasticity have been extensively studied in the literature \citep{Doering2005,Melbourne2008,Ovaskainen2010}. One of the most fundamental forms of stochasticity is demographic, where the probability of a birth (or death) occurring within a given time interval is drawn from a probability distribution, rather than simply occurring at a predetermined rate. The carrying capacity in this context is then defined as the number of individuals at which births and deaths are equal to each other in the mean-field limit (when the number of individuals is large). In general, the mean time to extinction of a population experiencing demographic stochasticity alone increases exponentially with $K$ \citep{Leigh1981,Lande1993,Doering2005}, meaning that large populations only rarely go extinct due to demographic stochasticity alone.

A second form of stochasticity is that due to a varying environment. Environmental stochasticity may take a vast array of different forms, giving rise to a similarly diverse array of mathematical approaches \citep{Ovaskainen2010}. Perhaps the most general model is to suppose that environmental stochasticity causes the population size to vary according to white or Ornstein-Uhlenbeck noise. As opposed to an exponential relationship, the mean time to extinction here increases roughly geometrically with $K$ \citep{Leigh1981,Doering2005}. 

Whereas an Ornstein-Uhlenbeck process is mathematically tractable in the case of large populations, it is more difficult to analyse in a birth-death model with discrete numbers of individuals. Furthermore, its influence upon population numbers is less physically intuitive than some other forms of environmental stochasticity. One more physical prescription is the ``catastrophe" model \citep{Mangel1993}, whereby catastrophes arrive at random intervals and remove a probabilistically-determined fraction of the existing population. Though intuitively appealing, such a catastrophe model suffers from the unrealistic assumption that catastrophes lasts an infinitesimally short amount of time, removing information regarding the population's trajectory shortly before extinction.

In this work, we investigate environmental stochasticity that lies somewhere in between the white noise and the catastrophe limits. Specifically, we suppose that the environment switches randomly between two states, a ``good" state, and a ``bad state," where the latter is defined as having an enhanced death rate. Each state lasts a length of time that is drawn from an exponential distribution, i.e., the switching constitutes a Telegraph Process \citep{Gardiner1985}. We note that similar scenarios have been considered previously in terms of the impact of a single bad event \citep{Assaf2009} (where instead of an enhanced death rate, a drop in birth rate was used), or under the assumption of a stationary probability distribution \citep{Hufton2016}.


Our physical set-up is reminiscent of a classical problem in physics - Brownian diffusion of a particle within a potential well \citep{Kramers1940,Doering1992}. Given enough time, the particle will eventually escape its potential well, where the mean time of escape may be computed using standard methods of stochastic calculus \citep{Gardiner1985,Mangel2006}. In this work, the population size represents the particle's position and the potential well represents the mean drift of the population (i.e., its evolution in the mean-field limit). Stochastic fluctuations in death rate between good and bad states are analogous to switching the depth and location of the potential well. 

Despite its simplicity, an analysis of the particle problem led to the discovery of a new phenomenon - ``resonant activation" - whereby the mean time of escape is minimized for a particular barrier fluctuation rate \citep{Doering1992}. Follow-up work sought to determine the generality of resonant activation \citep{Broeck1993,Iwan2003} and numerous studies have resurrected the conceptual result. For example it has been demonstrated that there exists an optimum migration rate between isolated biological populations that maximizes the mean extinction time of the metapopulation \citep{Khasin2012}. Within the context of cell biology, noise of a critical autocorrelation time minimizes the mean time of switching between two cellular phenotypes \citep{Assaf2013}. In addition to its theoretical robustness, resonant activation has been demonstrated experimentally \citep{Mantegna2000,Chizhevsky2009}.

A key finding we will present in this work is that our environmental fluctuation model displays resonant activation and that the model set-up facilitates an heuristic explanation of the process that the authors are not aware has appeared in the literature previously (but see ref. \citep{Khasin2012} for an explanation within the context of the migration model mentioned above). In what follows, we describe our methods for computing the extinction times and provide an heuristic explanation for the results, before briefly discussing implications of our findings.

  \begin{figure*}[ht!]
\centering
\includegraphics[trim=0cm 0cm 0cm 0cm, clip=true,width=1\textwidth]{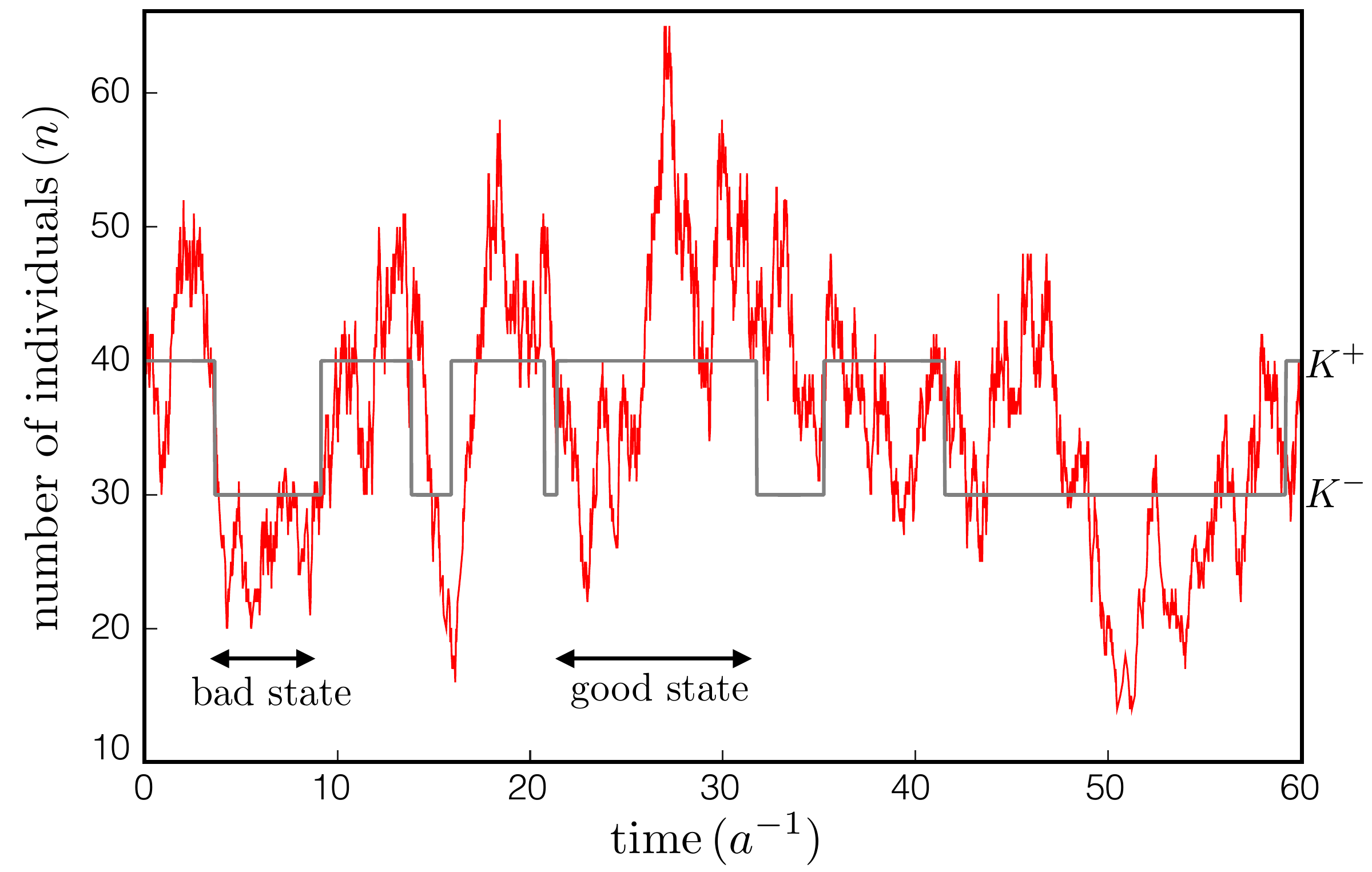}
\caption{A typical realization of the population size as a function of time (red line), computed using direct simulations as described in the text. The environment undergoes stochastic switching between two states (grey line) - the ``good" state with carrying capacity of $K^+=40$ and the bad state possesses $K^-=30$. For illustration, we chose parameters $\alpha=a/5$, $\epsilon=1$, $r_0=1/3$ and $A=1.2$. }
 \label{Schematic}
\end{figure*}

\section{Model Description}

Throughout this work we treat the population as a stochastic birth-death process, utilizing the Verhulst population model \citep{Nasell2001,Doering2005} with density-dependent death rates. Similar conclusions may be expected within the framework of other models, such as the SIS model. Our analytical methods will require the introduction of a population ceiling, which is defined as the population size $n=N_{\textrm{max}}$ at which the birth rate vanishes \citep{Mangel2006,Cairns2007}. The birth rate $\beta_n$ and death rate $\delta_n$ take the functional forms

\begin{align}
\beta_n&=
\begin{cases}
      an, & \text{if}\ n<N_{\textrm{max}} \\
      0, & \text{if}\ n=N_{\textrm{max}}
    \end{cases}\,\,\,\,\nonumber\\
\delta_n&=arn\bigg(1+\frac{n}{n'}\bigg)
\end{align}
where $a$ is the per capita birth rate (and $1/a$ may be conceptualized as a characteristic generation turnover timescale), $r$ is the low-density ratio of death rate to birth rate and $n'$ parameterizes the degree of density-dependence. In the mean field limit with a steady environment, the above birth and death rates yield an ODE for the number of individuals $N$
\begin{align}
\frac{dN}{dt}&=aN(1-r)\bigg(1-\frac{N}{K}\bigg),
\end{align}
where $K$ is typically referred to as the carrying capacity, and constitutes the equilibrium, or stationary, population size in the mean field limit (for $r<1$). Its value is related to $n'$ by 
\begin{align}\label{capacity}
K=\frac{(1-r)}{r}n'.
\end{align}

In the model we study here, the death rate is allowed to stochastically fluctuate between a high (``bad") state and a low (``good") state. Specifically, the value of $r$ in the definition of $\delta_n$ above evolves as a Telegraph Process \citep{Gardiner1985} taking one of two values
\begin{align}\label{states}
r&\rightarrow r_0, \,\,&\textrm{``good"\,\,state}\nonumber\\
r&\rightarrow A r_0, \,\,&\textrm{``bad"\,\,state}.
\end{align}
Such environmental switching may be thought of as a stochastic fluctuation between states with two different carrying capacities, the good state with $K^+$ and the bad state with $K^-$, following equation~\ref{capacity}. In all cases discussed below, we will maintain the same parameters for the good state, choosing $n'=20$ and $K^+=40$, corresponding to $r_0=1/3$. Given a carrying capacity in the good state of 40 individuals, a sufficiently high population ceiling is chosen as $N_{\textrm{max}}=100$, which we verify using numerical simulations (see for example Figure~(\ref{Schematic})), for which a population ceiling need not be specified.

The switching rate from the good state to the bad state is denoted $\alpha^+\equiv\alpha\epsilon$, and the rate of switching from bad to good is $\alpha^-\equiv \alpha$, where the case $\epsilon\ll1$ corresponds to a population subject to the influence of short-lived, catastrophic events. In this work, we only consider cases where $\epsilon\leq1$, and will vary $A,\,\epsilon$ and $\alpha$ in our analysis below.

\section{Computing mean extinction times}

\subsection{Fluctuating Environment}

Our goal in this work is to compute the mean time to extinction $\mathcal{T}_{n}$ for a population of $n$ individuals as a function of the environmental switching parameters. We will mostly restrict attention to scenarios beginning in the good state, the extinction time in which case being denoted with a ``+" in the superscript $\mathcal{T}^+_n$. However, the extinction time may be computed without knowledge of the initial environmental state ($\bar{\mathcal{T}}_n$) by marginalizing over the fluctuations, such that
\begin{align}\label{Margin}
\bar{\mathcal{T}}_n=\frac{\mathcal{T}_n^++\epsilon\,\mathcal{T}_n^-}{1+\epsilon}.
\end{align}
For the cases examined in this work ($\mathcal{T}_n^+\gg \mathcal{T}_n^-$ and $\epsilon\leq1$) $\bar{\mathcal{T}}_n$ behaves similarly to $\mathcal{T}^+_n$ (see Figure~\ref{EpsilonOne}) and so our conclusions regarding the latter inform the former. However, it is worth noting that $\mathcal{T}^-_n$ exhibits qualitatively different behaviour than $\mathcal{T}^+_n$, discussed briefly below.

 There exist multiple techniques for computing $\mathcal{T}_{n}^+$ (see e.g. \citep{Macarthur1967,Gardiner1985,Doering2005,Mangel2006}). We choose the method of averaging along immediate sample paths. Specifically, suppose that the population has $n$ individuals and the environment is in the good state. The mean time to extinction is then equal to the expected time before an event occurs, plus the mean of the extinction times after any one of the three possible events occurs (i.e., birth, death or environmental switch). Specifically, 
\begin{align}
\mathcal{T}_n^+&=\frac{1}{\beta_n^++\delta_n^++\alpha^+}+\frac{\beta_n^+}{\beta_n^++\delta_n^++\alpha^+}\mathcal{T}_{n+1}^+\nonumber\\
&+\frac{\delta_n^+}{\beta_n^++\delta_n^++\alpha^+}\mathcal{T}_{n-1}^++\frac{\alpha^+}{\beta_n^++\delta_n^++\alpha^+}\mathcal{T}_{n}^-\nonumber\\
\mathcal{T}_n^-&=\frac{1}{\beta_n^-+\delta_n^-+\alpha^-}+\frac{\beta_n^-}{\beta_n^-+\delta_n^-+\alpha^-}\mathcal{T}_{n+1}^-\nonumber\\
&+\frac{\delta_n^-}{\beta_n^-+\delta_n^-+\alpha^-}\mathcal{T}_{n-1}^-+\frac{\alpha^-}{\beta_n^-+\delta_n^-+\alpha^-}\mathcal{T}_{n}^+
\end{align}
which upon rearranging, become the governing equations
\begin{align}\label{meanTime}
-1=&\beta_n^+\mathcal{T}_{n+1}^++\delta_n^+\mathcal{T}_{n-1}^++\mathcal{T}_n^-\alpha^+\nonumber\\
&-\mathcal{T}_n^+(\beta_n^++\delta_n^++\alpha^+)\nonumber\\
-1=&\beta_n^-\mathcal{T}_{n+1}^-+\delta_n^-\mathcal{T}_{n-1}^-+\mathcal{T}_n^+\alpha^-\nonumber\\
&-\mathcal{T}_n^-(\beta_n^-+\delta_n^-+\alpha^-).
\end{align}
We may write equations~\ref{meanTime} as one matrix equation
\begin{align}\label{Governing}
-\mathbf{1}=\mathbf{M}\mathbf{T},
\end{align}
where the first $N_{\textrm{max}}$ elements of the vector $\mathbf{T}$ consist of $\mathcal{T}_n^+$ and the rest consist of $\mathcal{T}_n^-$. The above matrix equation is to be solved subject to the boundary conditions $\mathcal{T}_0^{+/-}=0$ and $\beta_{N_{\textrm{max}}}=0$.

\subsection{Static Environment}

A vast literature exists pertaining to the analysis of extinction times in static environments \citep{Doering2005}. We take advantage of this analytical understanding by expressing extinction times within the switching environment in terms of extinction times corresponding to 3 well-defined constant environmental states. These 3 states consist of the bad state, (mean extinction time $\tau^-_n$), the good state ($\tau^+_n$) and the ``mean" state ($\bar{\tau}_n$), the latter of which we define as the extinction time in the limit where $\alpha\rightarrow\infty$. In the mean state, the population evolves as if it were subject to time-independent ``mean" death and birth rates (derived below).

Using similar arguments as in the fluctuating environment, one can show that in the static environment,
\begin{align}\label{Paths}
-1=\beta_n\tau_{n+1}+\delta_n\tau_{n-1}-\tau_{n}(\beta_n+\delta_n).
\end{align}
where we obtain the equations for any one of the 3 static cases by adding superscripts $`+'$, $`-'$ or an overbar. Once again rewriting in matrix form we have that
\begin{align}\label{Static}
-\mathbf{1}&=\mathbf{M}^+\mathbf{\tau^+}\nonumber\\
-\mathbf{1}&=\mathbf{M}^-\mathbf{\tau^-}\nonumber\\
-\mathbf{1}&=\bar{\mathbf{M}}\,\,\mathbf{\bar{\tau}},
\end{align}
where each of $\mathbf{M}^+$, $\mathbf{M}^-$ and $\bar{\mathbf{M}}$ denote $N_{\textrm{max}}$ by $N_{\textrm{max}}$ matrices encoding the birth and death rates in the good, bad and mean states respectively.

In order to compute the appropriate birth and death rates for the mean state, we take the limit where $\alpha\rightarrow\infty$ in equations~\ref{meanTime}, from which we extract that 
  \begin{align}
  \mathcal{T}^+_{n}\big|_{\alpha\rightarrow\infty}\approx \mathcal{T}^-_{n}\big|_{\alpha\rightarrow\infty}=\bar{\tau}_n,
  \end{align}
 and, upon convert Equation~\ref{meanTime} into a form similar to expression~\ref{Paths},
  \begin{align}
-1=\bar{\beta}_n\bar{\tau}_{n+1}+\bar{\delta}_n\bar{\tau}_{n-1}-\bar{\tau}_{n}(\bar{\beta}_n+\bar{\delta}_n),
  \end{align}
  where we arrive at the mean birth and death rates 
  \begin{align}
  \bar{\beta}_n&\equiv \frac{\beta_n^++\epsilon \beta_n^-}{1+\epsilon}=an\nonumber\\
  \bar{\delta}_n&\equiv \frac{\delta_n^++\epsilon \delta_n^-}{1+\epsilon}=an\,r_0\bar{A}\bigg(1+\frac{n}{n'}\bigg).
  \end{align}
  The mean value for the environmental parameter is
\begin{align}\label{meanA}
\bar{A}\equiv \frac{1+\epsilon A}{1+\epsilon}.
\end{align}
If we consider the mean-field evolution of a population in the regime where $\alpha\rightarrow\infty$, it will possess an stationary number of individuals, or carrying capacity given by
\begin{align}\label{meanK}
K^\infty\equiv\frac{1-r_0\bar{A}}{r_0\bar{A}}n'.
\end{align}

It should be noted that all extinction times computed above are sensitive to the number of individuals. However the ratio $T_{N_{max}}/T_1$ is never more than a factor of $\sim2-4$ (at least in the cases considered here). This can be seen most readily with respect to the static environments. In particular, suppose that we are only interested in cases where $\tau^{+/-}\gg 1/(\beta_n^{+/-}+\delta_n^{+/-})$ (i.e., $r<1$) and $n$ small enough such that the death rate is linear $n\ll n'$. In this case equation~\ref{Paths} becomes
\begin{align}
\tau_n^{+/-}\approx\frac{1}{1+r}\tau_{n+1}^{+/-}+\frac{r}{1+r}\tau_{n-1}^{+/-},
\end{align}
which may be solved to obtain
\begin{align}\label{recursion}
\tau_n^{+/-}\approx\tau^{+/-}_{N_\textrm{max}}(1-r^n),
\end{align}
an expression which is equivalent to that derived in the large population limit in reference~\citep{Doering2005} (their eq.~(20)). The factor $\tau^+_{N_{max}}/\tau^+_1\approx1/(1-r)=3/2$, where we have used the value $r=r_0=1/3$ for the ``good state". In the bad state, $r=r_0A$ which may exceed unity, suggesting a negative extinction time by equation~\ref{recursion}, however this would break the initial assumption that the extinction time is significantly longer than the typical time between events. 

The dependence of extinction time upon the initial population size may be important when one considers situations where populations are initiated at very low densities of individuals, such as island colonization \citep{Macarthur1967}, however it is not crucial to consider here. When quoting the extinction times throughout the paper, unless otherwise stated, we choose $n$ to be the carrying capacity in the good state ($n=K^+=40$). Accordingly, for ease of presentation, we define 
 \begin{align}
 \tau^+&\equiv \tau^{+}_{K^+}\nonumber\\
 \tau^-&\equiv \tau^{-}_{K^+}.
 \end{align}
 The calculations outlined in this section introduce 3 states that are simpler to understand, both conceptually and mathematically, than the full fluctuating problem.
 \subsection{Direct Simulation}
The time evolution of the population number $n$ may be computed numerically in order to check the solutions obtained through our analytic methods. Specifically, at each time step, two random numbers are drawn, $R_1$ and $R_2$ each uniformly distributed between 0 and 1. The first random number $R_1$ is used to determine the wait time before an event occurring
\begin{align}
\Delta t=-\frac{\ln R_1}{\gamma^{+/-}},
\end{align}
where we define the sum of rates 
\begin{align}
\gamma^{+/-}\equiv\beta_n^{+/-}+\beta_n^{+/-}+\alpha^{+/-}.
\end{align}
The second random number $R_2$ is used to determine which event occurs (birth, death or environment switch), following the prescription:
\begin{align}
&n^{+/-}\rightarrow (n+1)^{+/-}\,\,\,\textrm{if}\,\,\,R_1<\beta_n^{+/-}/\gamma^{+/-}\nonumber\\
&n^{+/-}\rightarrow (n-1)^{+/-}\,\,\,\textrm{if}\,\,\,\nonumber\\
&R_1>\beta_n^{+/-}/\beta\,\,\&\,\,R_1<(\beta_n^{+/-}+\delta_n^{+/-})/\gamma^{+/-}\nonumber\\
&n^{+/-}\rightarrow n^{-/+}\,\,\,\textrm{if}\,\,\,R_1>(\beta_n^{+/-}+\delta_n^{+/-})/\gamma^{+/-}.
\end{align}
Each time we quote a mean extinction time using direct simulation, we use the above algorithm to compute 100 trajectories, beginning with $n=K^+=40$ individuals in the good state, and average the extinction times. For illustrative purposes, a typical realization of the population number is presented in Figure~\ref{Schematic}, with $\alpha=a/5$, $\epsilon=1$ and $A=1.2$. Note that the direct simulation does not require a population ceiling, and our choice of ceiling at $n=N_{\textrm{max}}=100$ in the analytic techniques was informed by the rarity with which $n$ reaches 100 individuals. 

\section{Results \& Analysis}
\subsection{Case where $\alpha^+=\alpha^-$}

The first case we explore is a system that spends, on average, equal amounts of time in the good state as the bad state ($\alpha^+=\alpha^-$; $\epsilon=1$). Let us suppose that in the good state, the carrying capacity $K(\equiv K^+)=40$ but it drops to $K(\equiv K^-)=30$ in the bad state. Using equations~(\ref{capacity},\,\ref{states}), this may be modelled by choosing $A=1.2$, $r_0=1/3$ and $n'=20$. With these numerical values, the mean state ($\alpha/a\rightarrow\infty$) is described with $\bar{A}=1.1$ and a carrying capacity of $K^\infty=34.5$ (equs~(\ref{meanA},\ref{meanK})). 

 \begin{figure}[ht!]
\centering
\includegraphics[trim=0cm 0cm 0cm 0cm, clip=true,width=1\columnwidth]{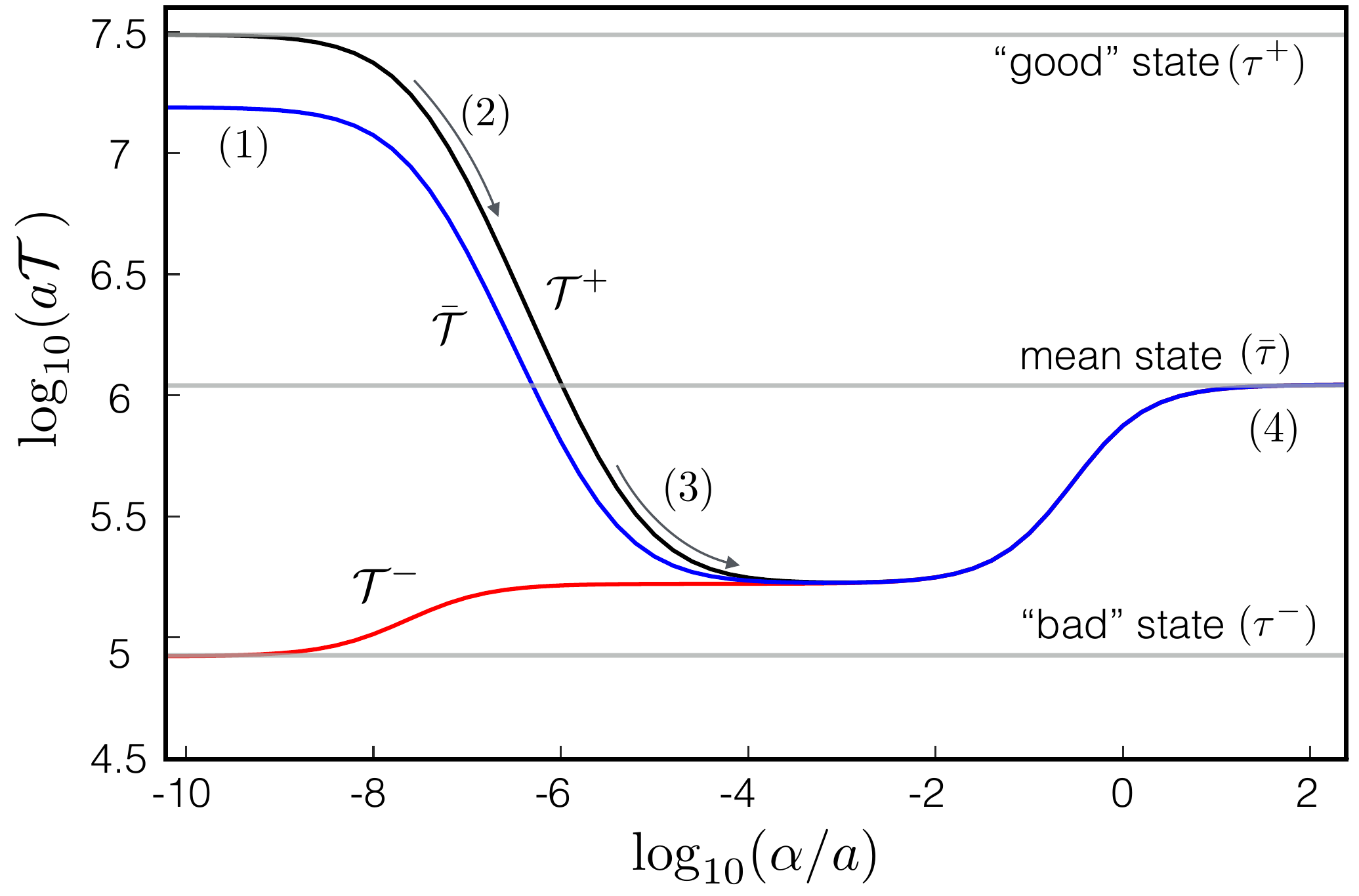}
\caption{The mean extinction time as function of environmental switching rate. The black curve correspond to an initial condition of the good state, the red curve corresponds to beginning in the bad state and the blue curve represents the extinction time averaged over both initial conditions following equation~\ref{Margin}. Here, $\epsilon=1$, $n'=20$ and the carrying capacities in the good and bad states are $K^+=40$ and $K^-=30$ respectively. We denote, with horizontal lines, the mean extinction times ($\tau^+$ and $\tau^-$; computed using Equation~\ref{Paths}) corresponding to these two states, along with a third line representing the extinction time in the mean state $\bar{\tau}$. The numbers in parentheses are referred to in the text.}
 \label{EpsilonOne}
\end{figure}

In Figure~\ref{EpsilonOne}, we illustrate the mean extinction times beginning in the good state $\mathcal{T}^+$, the bad state $\mathcal{T}^-$ and the average of these two $\bar{\mathcal{T}}$ as a function of the environmental switching parameter $\alpha$, where all times are computed with $n=K^+$. For small $\alpha/a$, all three curves are flat, with $\mathcal{T}^+\approx\tau^+$ and $\mathcal{T}^-\approx\tau^-$. When the population begins in the bad state, the extinction time rises monotonically with $\alpha$ (red curve), essentially owing to the fact that as $\alpha$ is increased, the system beginning in the bad state becomes more likely to survive into the next good state, lengthening its persistence. In contrast, the curve of $\mathcal{T}^+$ falls initially, but begins to rise again after reaching a minimum value. As expected, all curves approach $\bar{\tau}$ in the limit where $\alpha/a\rightarrow\infty$.

In this work, we are most interested in the occurrence of a minimum in the extinction curves of $\bar{\mathcal{T}}$ and $\mathcal{T}^+$. Furthermore, owing to our consideration of $\epsilon\leq1$, we do not discuss in detail the curve of $\mathcal{T}^-$. However the logic acquired from our discussion of $\mathcal{T}^+$ may be easily applied to $\mathcal{T}^-$. 

 The existence of a minimum in the extinction time is a signature of resonant activation \citep{Doering1992}. For the parameters chosen here, the minimum remains flat across roughly 3 orders of magnitude in $\alpha/a$ from $-4\lesssim \log_{10}(\alpha/a)\lesssim-1$. This result says that the population will typically go extinct fastest when subject to environmental perturbations acting once every 10-10,000 generations.

 \subsection{Heuristic Explanation}

Multiple previous scenarios demonstrating resonant activation have been studied, but the heuristic mechanism responsible has often not been identified. In this section, we discuss the mechanism from a qualitative point of view in order to better understand why resonant activation occurs in this system.

 Beginning in the limit $\alpha/a\rightarrow0$, both states (good and bad) last such a long time that the extinction time in the fluctuating environment approaches that of the initial state, i.e. $\tau^+$ if the system begins in a good state and $\tau^-$ if it begins in the bad state (``1" in Figure~\ref{EpsilonOne}). As $\alpha$ is increased, populations beginning in the good state will typically experience an environmental switch before going extinct, but the following bad state still lasts long enough to almost ensure extinction. When $\alpha$ enters this regime (at $\alpha\epsilon\approx1/\tau^+=10^{-7.5}a$) the graph steepens to a slope of roughy $1/\alpha\epsilon$ (``2" in Figure~\ref{EpsilonOne}).

Eventually, as $\alpha$ increases further, the mean extinction time in the bad state ($\tau^-$) becomes longer than the average duration of a single bad state ($1/\alpha$). Consequently, the population gets ``saved" before going extinct as the environment switches back to the good state. At this point, the extinction time in the fluctuating problem flattens again (``3" in Figure~\ref{EpsilonOne}). From figure~\ref{EpsilonOne}, we see that the bad state has a mean extinction time of $\tau^-\approx10^{4.9}/a$ and so the turnoff into a minimum occurs at the expected $(\tau^-)^{-1}=\log_{10}(\alpha/a)\approx-4.9$.

\begin{figure*}[ht!]
\centering
\includegraphics[trim=0cm 0cm 0cm 0cm, clip=true,width=1\textwidth]{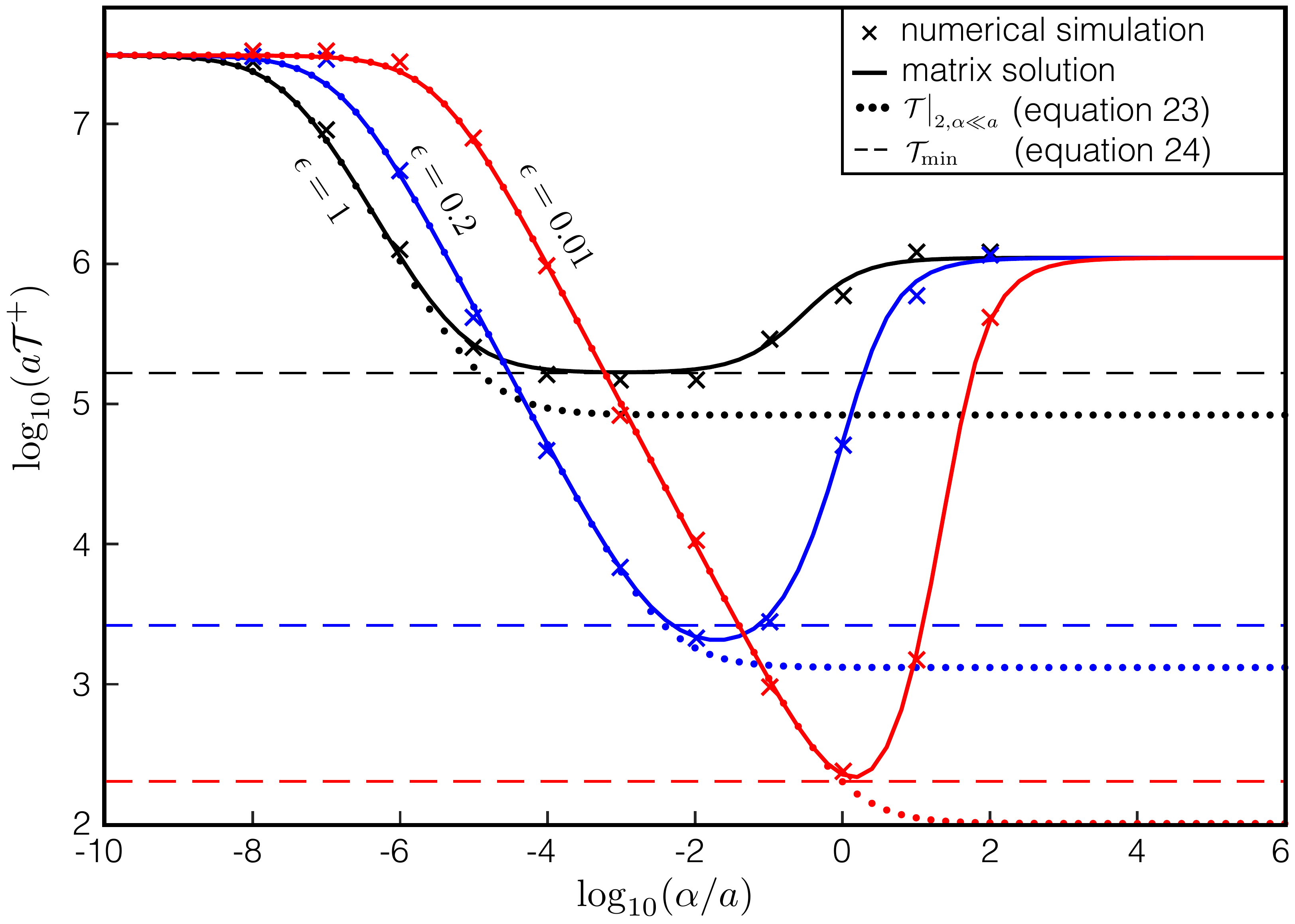}
\caption{The mean extinction time as a function of $\alpha$ for 3 different choices of $\epsilon$=(1, black; 0.2, blue; 0.01, red), all possessing equivalent $\alpha\rightarrow\infty$ extinction times ($\bar{\tau}$). Four metrics are plotted: crosses denote the mean extinction time computed using direct simulations, solid lines were computed using the matrix equation~\ref{Governing}, dotted lines depict our heursitcally-derived approximation in the regime where $\alpha\ll a$ (equation~\ref{LowalphaApprox}) and the dashed line denotes the expected minimum extinction time (equation~\ref{Minimum}) using the arguments in the text. As expected, red and blue graphs turn off into the minimum at of $1/\alpha$ roughly corresponding to the extinction time in the bad state $\log_{10}(\alpha/a)=-2.4$ for $A=1.6$ and $\log_{10}(\alpha/a)=0$ for $A=11.1$.}
\label{ConstantMean}
\end{figure*}
In the limit where $\alpha\rightarrow\infty$, the mean extinction time becomes equivalent to that of a state where the death rate is described by the mean value of $A$ (``4" in Figure~\ref{EpsilonOne}). For the parameters chosen here, the mean state is characterized by a mean extinction time $\bar{\tau}$ which is greater than the magnitude of $\mathcal{T}^+$ at the point where $\alpha\approx1/\tau^-$ and thus the minimum exists and resonant activation is observed. Later on, we discuss the criteria under which the graph does not display a minimum.
\begin{figure*}
\centering
\includegraphics[trim=0cm 0cm 0cm 0cm, clip=true,width=1\textwidth]{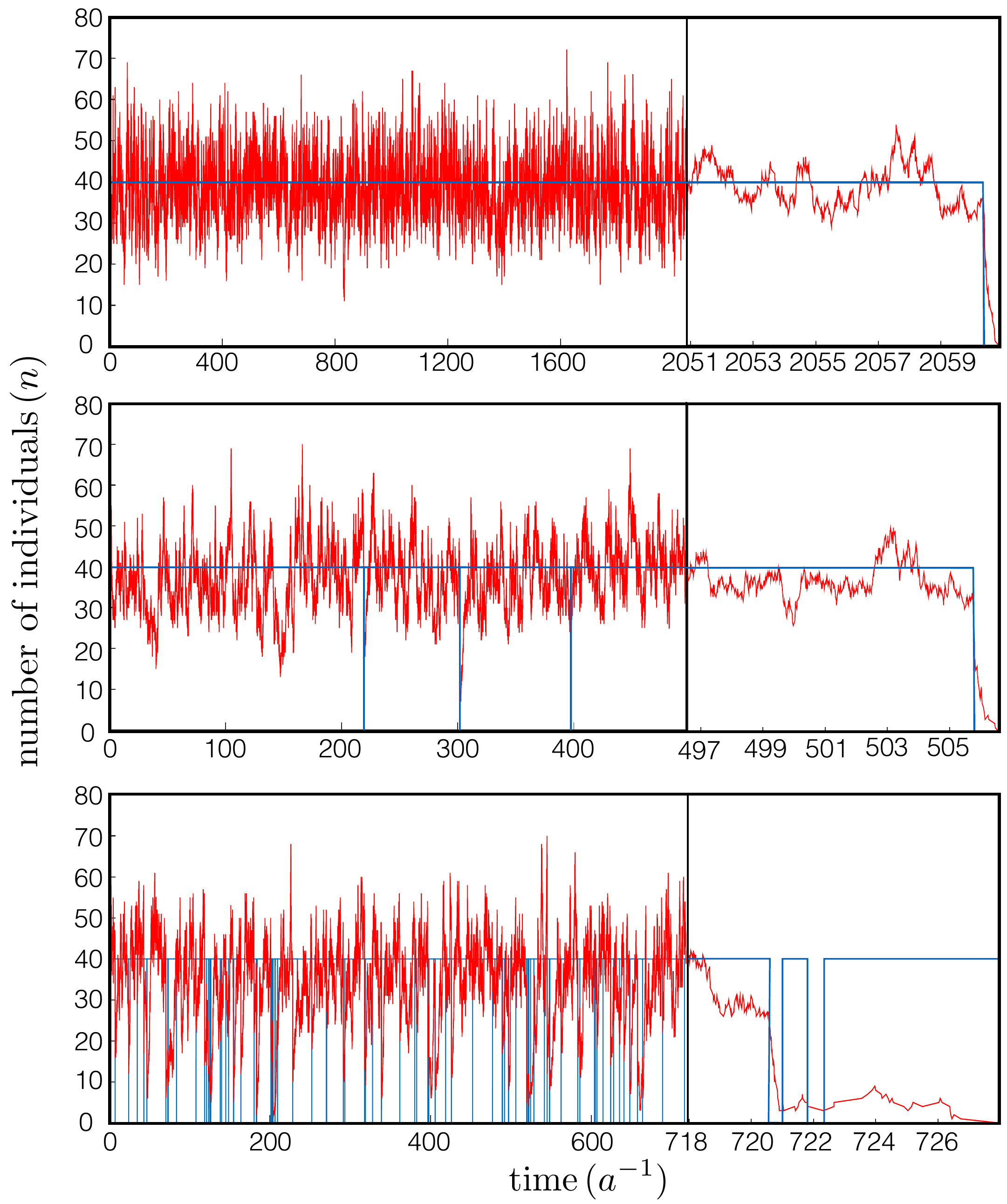}
\caption{Typical realizations of populations going extinct for 3 values of $\alpha=0.1a$ (top panel), $a$ (middle panel) and $10a$ (bottom panel). All cases have $\epsilon=0.01$ and possess bad events with $A=11.1$, for which $K^-=-14.6<0$, such that the blue curve, reflecting the time evolution of the carrying capacities, transitions from $K^+=40$ to the below the $y$-axis each time a bad event occurs. These parameters are the same as those used to generate the red curve in Figure~\ref{ConstantMean} and the 3 values of $\alpha$ are chosen to lie below, close to and above the value of $\alpha$ at which the minimum extinction time occurs, in order to illustrate the difference in the path to extinction within the 3 cases.}
\label{Regimes}
\end{figure*}

\subsection{Catastrophe case}

In the previous section we considered a case where the system spends on average equal amounts of time in the two different states, where the states differ in carrying capacity by 10. We now examine a more general scenario. Specifically, the bad state lasts a shorter time on average than the good state, but its severity $A$ is increased such as to maintain a constant mean state $\bar{A}=1.1$, allowing comparison to the previous section. 

The situation thus described is more similar to a catastrophic event \citep{Lande1993,Assaf2009}. In Figure~\ref{ConstantMean}, we compare the mean extinction times for 3 different values of $A$ and $\epsilon$ that maintain equivalent $\bar{A}$. Specifically, we choose $A=1.2$, $\epsilon=1$ as before, but include two examples of catastrophes; $A=1.6$, $\epsilon=0.2$, and $A=11.1$, $\epsilon=0.01$. Once again a minimum is observed in these two additional cases, though their minima are sharper than for $\epsilon=1$.

In addition, we perform direct numerical simulations at multiple values of $\alpha$, where as mentioned above we average the extinction times of 100 paths. The resulting mean extinction times are plotted as crosses on Figure~\ref{ConstantMean}. The excellent agreement between the solid lines and crosses validates both our matrix solution (equation~\ref{Governing}) and the choice of $N_{\textrm{max}}=100$.

Using the intuition gained from the previous section, we would expect the mean extinction time curves to change slope at $\alpha\epsilon\approx 1/\tau^+$ and then again at $\alpha\approx 1/\tau^-$. For $\epsilon=0.2$, these two values correspond to $\log_{10}(\alpha/a)=\{-6.8,\,-2.4\}$, and for $\epsilon=0.01$, they correspond to $\log_{10}(\alpha/a)=\{-5.5,\,0\}$. By inspection of Figure~\ref{ConstantMean}, we see that the turning points match well with these values, validating an equivalent physical interpretation between the catastrophe and equal-time cases.


For illustration, in Figure~\ref{Regimes}, we provide example trajectories for populations going extinct under switching rates near the minimum of the $\epsilon=0.01$ curve and on either side of the minimum. Specifically, at $\alpha=0.1a$ (top panel) the bad events last long enough to lead to extinction most of the time and so the curve drops to zero individuals soon after the environment switches. At $\alpha=a$ (middle panel) the system exhibits its minimum extinction time, where the catastrophic events are occurring both frequently and with a significant chance of extinction. For more frequent catastrophes $\alpha=10a$ (bottom panel) the bad events are more frequent, but too short to lead to extinction in most cases. Here extinction is caused by chance clustering of bad events and/or negative excursions in the good state.
\begin{figure*}[ht!]
\centering
\includegraphics[trim=0cm 0cm 0cm 0cm, clip=true,width=1\textwidth]{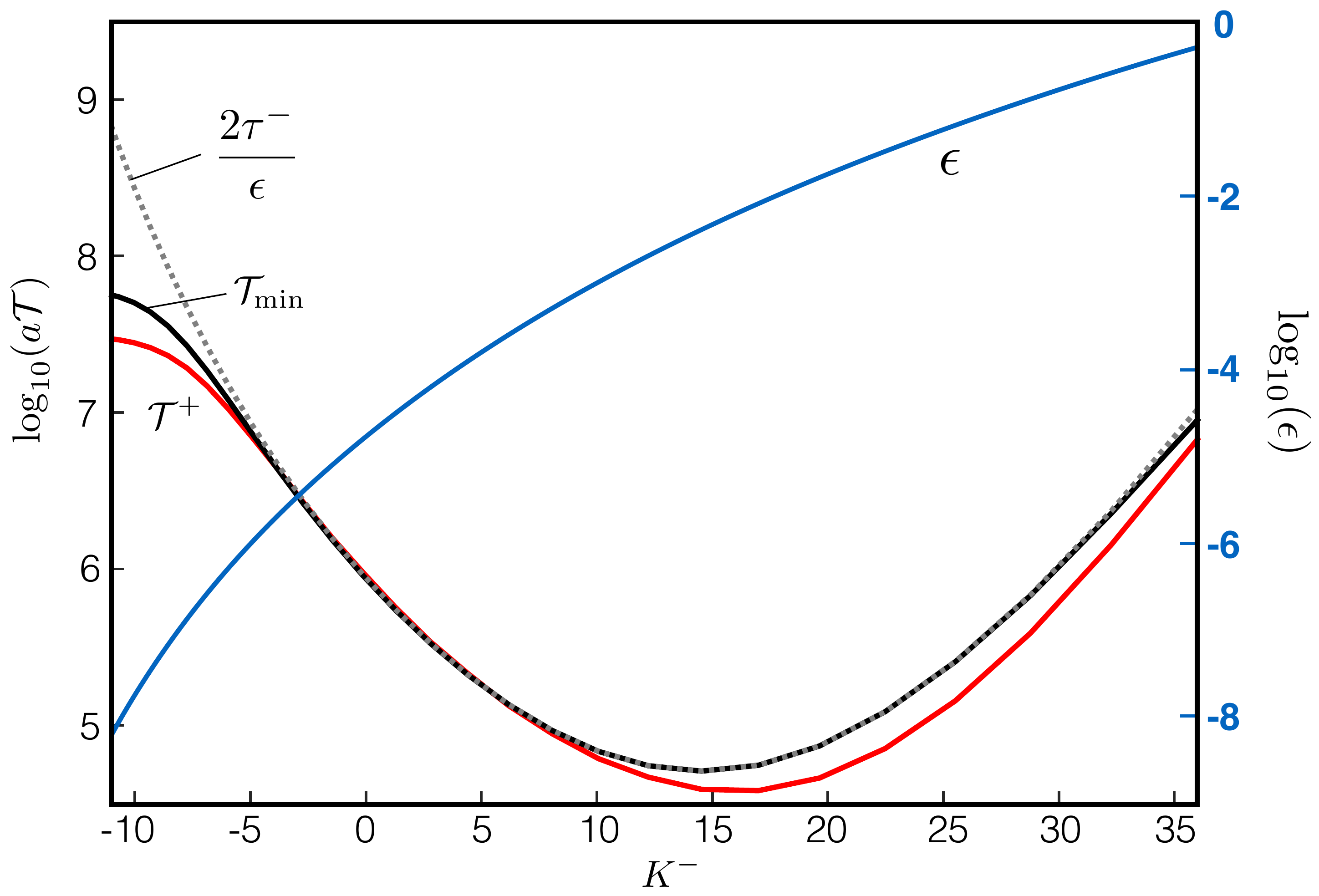}
\caption{The minimum time to extinction as a function of the carrying capacity in the bad state $K^-$, (solid red line) compared with approximation~\ref{Minimum} (solid black line) and its approximate form when $\epsilon \tau^+\gg\tau^-$ (dotted grey line). Both $A$ and $\epsilon$ (blue line) were varied (such that a minimum existed in all cases). We see that the approximation derived from heuristic arguments provides in general a slight over-estimate of the mean extinction time, but a close match overall.}
  \label{MinGraph}
\end{figure*}

\begin{figure*}[ht!]
\centering
\includegraphics[trim=0cm 0cm 0cm 0cm, clip=true,width=1\textwidth]{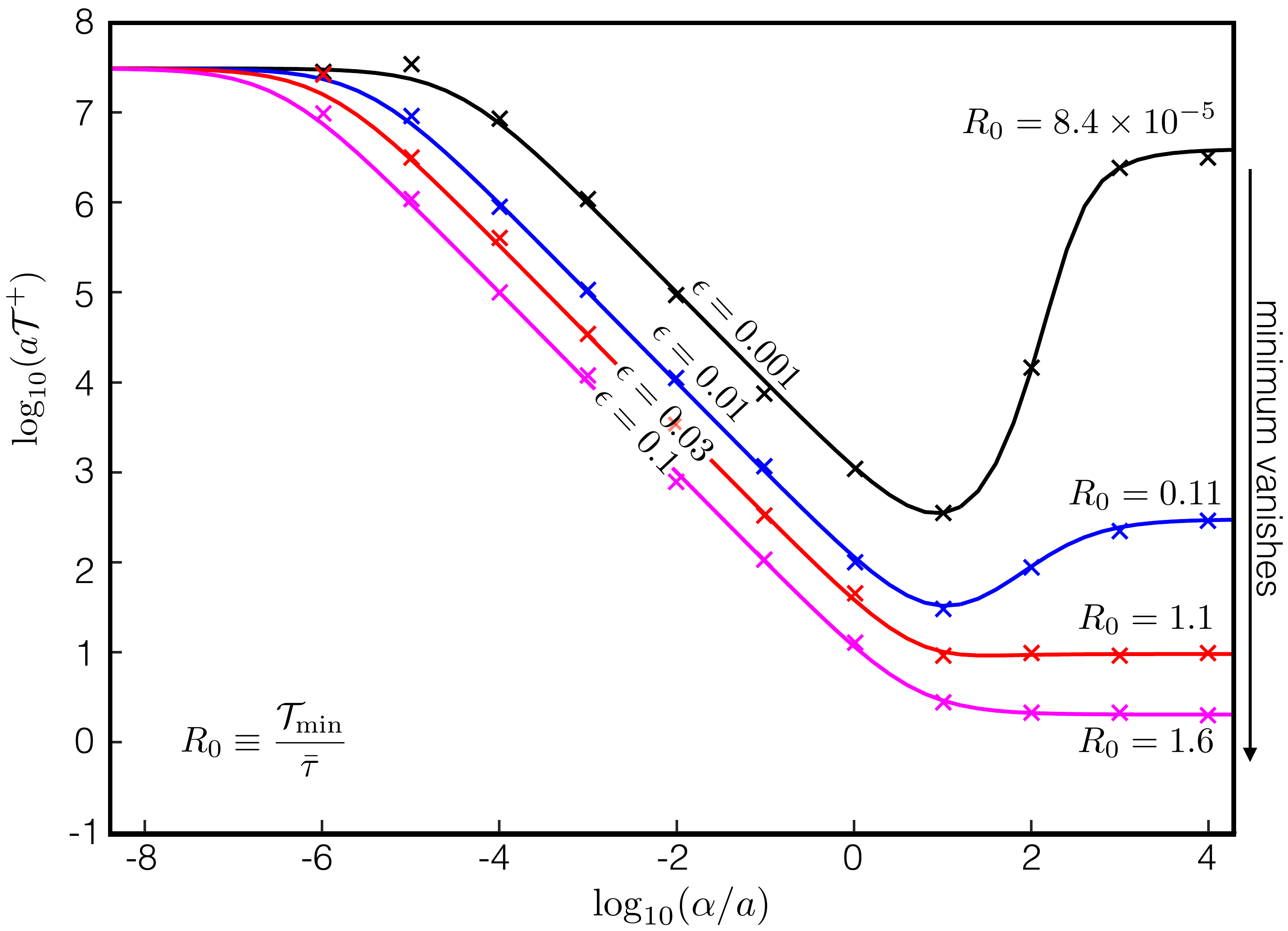}
\caption{An illustration of the disappearance of the minimum when the mean state is bad enough such that the parameter $R_0\equiv\ \mathcal{T}_{\textrm{min}}/\bar{\tau}>1$. All cases have $A=60$ and $n'=20$ but $\epsilon$ takes the values $\{0.001,\,0.01,\,0.03,\,0.1\}$. As $\epsilon$ is increased, the mean state worsens, removing the presence of a minimum. Crosses represent  extinction times derived from numerical simulations.}
  \label{noMin}
\end{figure*}
\subsection{Value of the minimum}
Given the regime changes in $\mathcal{T}^+$ as a function of $\alpha$ computed above, it is possible to derive an expression approximating the extinction time curve for $\alpha\ll a$. In particular, we want an expression that approaches $\tau^+$ for $\alpha\rightarrow 0$. Then, when $\alpha\epsilon\gtrsim (\tau^+)^{-1}$ the curve becomes approximately equal to the typical switching time. The two criteria above are matched by supposing that
\begin{align}\label{LowalphaApprox1}
\mathcal{T}\big|_{1,\alpha\ll a}\approx \tau^+\Bigg(\frac{1}{1+\alpha \epsilon \tau^+}\Bigg)
\end{align}
The approximation is improved by noting that as $\alpha$ increases further, $\mathcal{T}\big|_{1,\alpha\ll a}$ approaches zero and so must be corrected at larger $\alpha$. We match the turnover at $\alpha\approx(\tau^-)^{-1}$ by adding to the numerator $\alpha \tau^-$ such that a better approximation becomes 
\begin{align}\label{LowalphaApprox}
\mathcal{T}\big|_{2,\alpha\ll a}\approx \tau^+\Bigg(\frac{1+\alpha \tau^-}{1+\alpha \epsilon \tau^+}\Bigg)\,\,\,\,\,\,\,\,\,\,(\alpha\lesssim a).
\end{align}
We plot the above approximation in Figure~\ref{ConstantMean}, showing that, simply using the heuristic arguments outlined above, we approximate the curve well up until the minimum is encountered. More rigorous mathematical techniques may indeed yield an expression such as~\ref{LowalphaApprox}, but a detailed proof is beyond the scope of this work.

Given that expression~\ref{LowalphaApprox} closely matches the extinction time until the slope changes at $\alpha\sim1/\tau^+$ in the exact solution, the minimum extinction time for the switching problem $\mathcal{T}^+$ may be approximated by substituting $\alpha=1/\tau^-$ into the expression~\ref{LowalphaApprox}. The estimate thus obtained for the minimum extinction time is
\begin{align}\label{Minimum}
\mathcal{T}_{\textrm{min}}&\approx\frac{\,\,2 \tau^+\tau^-}{\epsilon \tau^++\tau^-}
\end{align}
We include horizontal lines on Figure~\ref{ConstantMean} at the extinction times predicted by this expression. Furthermore, we test the validity of approximation~\ref{Minimum} by plotting the true minimum extinction time against $K^-$ for a range of values of $\epsilon$ and $A$, alongside the approximation $\mathcal{T}_{\textrm{min}}$ (Figure~\ref{MinGraph}). We include a plot of $\mathcal{T}_{\textrm{min}}\approx 2\tau^-/\epsilon$, i.e., the limit of $\mathcal{T}_{\textrm{min}}$ where $\epsilon \tau^+\gg\tau^-$. Both approximations provide good estimates for the true minimum extinction time, though some disagreement arises owing to the qualitative nature of the derivation.

\subsection{Criterion for the existence of a minimum}
It is not guaranteed that expression~\ref{Minimum} for $\mathcal{T}_{\textrm{min}}$ will be less than the mean extinction time in the mean state. Accordingly the extinction time $\mathcal{T}^+$ will only exhibit resonant activation if the inequality
\begin{align}\label{MinCriterion}
\mathcal{T}_{\textrm{min}}&\lesssim\bar{\tau},
\end{align}
is satisfied. 
 In order to illustrate the above criterion, we compute the mean extinction times for 4 cases where $\epsilon$ is gradually increased (choosing $\epsilon=\{10^{-3},\,10^{-2},\,0.03,\,0.1\}$), thus decreasing $\bar{\tau}$ until no minimum is present in the extinction time curve. In order to remove the minimum, the mean state must exhibit a fairly short extinction time and so we define a bad state with $A=60$, chosen simply to ensure that a minimum exists for values of $\epsilon<1$. We illustrate these cases in Figure~\ref{noMin}.

 As can be seen, the high-$\alpha$ part of the graph drops and eventually the entire graph takes the form of a monotonic decline in mean extinction time from small to larger $\alpha$. For each case, we quote the ratio 
\begin{align}
R_0\equiv\frac{\mathcal{T}_{min}}{\bar{\tau}}
\end{align}
which we predict to be less than unity when a minimum exists, with that minimum occurring at $\alpha_{\textrm{min}}\sim1/T^-$. The numerical results plotted in Figure~\ref{noMin} agree well with our analytic expectations. 


\subsubsection{Large $\alpha$ expansion.}

Criterion~\ref{MinCriterion} provides a useful inequality for determining the existence of resonant activation based on qualitative arguments. However, one may obtain a more quantitative criterion for the existence of a minimum. Specifically, we perform an expansion of the system about large $\alpha$. If a minimum exists, then the gradient with respect to $\alpha$ at large $\alpha$ must be positive.
\begin{figure*}[ht!]
\centering
\includegraphics[trim=0cm 0cm 0cm 0cm, clip=true,width=1\textwidth]{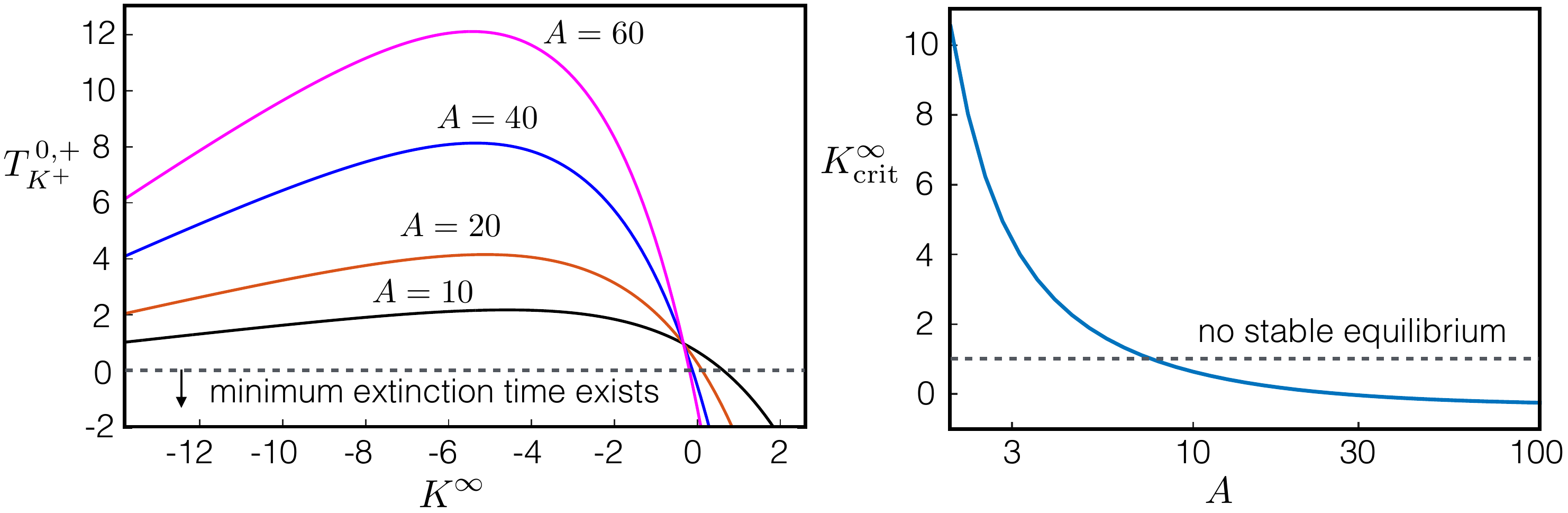}
\caption{Left: The first order correction term $T_{K^+}^{0,+}$ (Equ.~\ref{LargeAlpha}) to the extinction time as $\alpha\rightarrow\infty$. When $T_{K^+}^{0,+}<0$ the system exhibits resonant activation (i.e., a minimum extinction time). Right: The critical value of the carrying capacity in the mean state $K^{\infty}$, above which the system exhibits resonant activation as a function of $A$ (i.e., a minimum exists if $K^{\infty}>K^{\infty}_{\textrm{crit}}$). As the bad state gets worse ($A$ increases), the critical carrying capacity crosses $K^{\infty}_{\textrm{crit}}=1$, below which the mean state possess no stable stationary number of individuals in the mean-field limit.}
  \label{Kmin}
\end{figure*}

After some algebra, the solution to the matrix equation~\ref{Governing} may be written in closed form as 
\begin{align}
\mathbf{\mathcal{T}}^+&=\bigg[\mathbf{I}-\frac{1}{\alpha^++\alpha^-}\bar{\mathbf{M}}^{-1}\mathbf{M}^-\mathbf{M}^+\bigg]^{-1}\nonumber\\
&\times\bigg[\mathbf{I}-\frac{1}{\alpha^++\alpha^-}\bar{\mathbf{M}}^{-1}\mathbf{M}^-\bar{\mathbf{M}}\bigg]\bar{\mathbf{\tau}},
\end{align}
where the various matrices take on the forms introduced in equations~\ref{Static}. Expanding to first order in $a/\alpha$ we obtain the expression
\begin{align}
\mathbf{\mathcal{T}}^+&\approx\bigg[\mathbf{I}+\frac{1}{\alpha^++\alpha^-}\bar{\mathbf{M}}^{-1}\mathbf{M}^-(\mathbf{M}^+-\bar{\mathbf{M}})\bigg]\mathbf{\bar{\tau}}
\end{align}
Accordingly, the sign of the first order correction term 
\begin{align}\label{LargeAlpha}
\mathbf{T}^{+,0}\equiv\bigg[\bar{\mathbf{M}}^{-1}\mathbf{M}^-(\mathbf{M}^+-\bar{\mathbf{M}})\bigg]\mathbf{\bar{\tau}},
\end{align}
constitutes a test for the existence of resonant activation, where negative $\mathbf{T}^{+,0}$ corresponds to resonant activation occurring. 

In the left panel of Figure~\ref{Kmin} we compute $\mathbf{T}^{+,0}$ for four cases. In each case, $A$ is held fixed whilst $\epsilon$ varied. Upon plotting $\mathbf{T}^{+,0}$ against the resulting $K^{\infty}$, we see that $\mathbf{T}^{+,0}$ changes sign at values of $K^{\infty}\approx 1$, with the transition occurring closer to $K^{\infty}=0$ as $A$ is increased (Figure~\ref{Kmin}, right panel). Generally speaking, these results suggest that resonant activation will only occur if the mean state possess a sufficiently long extinction time (either by lowering $A$ or $\epsilon$).

\section{Discussion}
In this work, we have computed the mean time to extinction of a stochastic birth-death process subject to environmental forcing. Such forcing was modelled as a random fluctuation in the death rate, with characteristic frequency $\alpha$, between two states. We demonstrated, through both analytical techniques and numerical simulation that this system exhibits ``resonant activation" \citep{Doering1992} whereby there exists a fluctuation timescale that minimizes the mean time to extinction.

A key contribution of our work has been to provide a heuristic explanation for the emergence of resonant activation. However, it is not immediately obvious whether the results are specific to our chosen form of environmental forcing, i.e. where bad states typically last a time interval of $1/\alpha\epsilon$ and the good states last $1/\alpha$, with $\epsilon\leq1$. Environmental noise is unlikely to behave in this manner identically, but our framework ensures that rare events are more likely to cause extinction (because they last longer). At small $\alpha$ the bad states last long enough to almost ensure extinction and so the mean extinction time decreases with $\alpha$. When $1/\alpha\epsilon$ begins to exceed the mean extinction time within the bad state, more than one bad event is required for extinction and so the mean extinction time begins to rise again, hence the minimum in the extinction time curve.

Regardless of the exact form of forcing, the understanding acquired in our specific model may be applied to more general systems. For example reference~\citep{Assaf2013} examined a system forced by noise parameterized as an Ornstein-Uhlenbeck process with autocorrelation time, analogous to our fluctuation timescale $1/\alpha$. They too found that the mean escape time was minimized at a given autocorrelation time. Using our qualitative understanding we may similarly conclude that the minimum arises owing to the system reaching a point that is balancing the severity of rare events with the time interval between them.


We emphasize that all of our timescales are scaled by $1/a$, and so the minimum in extinction time should be interpreted as a minimum number of generations before extinction. A focus on generation number naturally leads to a discussion of how the picture might alter in the presence of biological evolution. In particular, we may consider two values of $\alpha$ sharing the same $\mathcal{T}^+$ (i.e., on either side of the minimum). As illustrated in Figure~\ref{Regimes}, the exact mechanism of extinction differs between the higher $\alpha$ and lower $\alpha$ regimes. At lower $\alpha$ extinction occurs during one long-lived bad event. In contrast, at larger $\alpha$ the population survives numerous short-lived bad events, with extinction occurring as a result of unfortunate clustering of events. 

With no evolution, the population typically goes extinct after the same number of generations in both cases. In reality the population in the faster switching environment will have encountered the bad state before, giving it the opportunity to adapt (essentially decreasing $A$) before finally going extinct. We thus speculate that evolutionary adaptation may extend the extinction time at $\alpha$ larger than the ``resonant" value, perhaps sharpening the minimum in real populations - species adapt more easily to the events that occur more frequently. The topic would benefit from future work that examines more generally the importance of adaptation within the context of resonant activation. 

Although our model has been conceptualized thus far as pertaining to births and deaths of individuals within a population, a potential consequence to evolution emerges if we apply it instead to specific alleles in a population. In particular, suppose a population possesses two alleles for one gene, but the species' fitness optimum sits at some non-zero ratio between the 2 alleles (a form of fitness landscape \citep{Wright1932,Niklas1994}). If the fitness optimum is fluctuating, then the mean time to fixation of one allele will depend upon the timescale of fluctuations. Resonant activation within this picture would manifest as a minimum fixation time of one of the alleles, essentially maximizing the evolutionary rate.
 
Though qualitative, the above argument suggests that stochastic timescales may have important influences upon evolutionary rates. Indeed, it is interested to highlight recent work demonstrating the importance of environmental fluctuation rate to fixation probabilities of one population over another \citep{Ashcroft2014,Melbinger2015}. This scenario is conceptually similar to the case of two alleles, lending credence to the potential for noise timescales to influence evolutionary rates. More work is needed to examine this further.
 
 A separate interpretation of our model, aside from births and deaths of individuals, is the originations and extinctions of different taxa within a lineage \citep{Macarthur1967,Sepkoski1984}. In this case the per-taxa speciation timescale $1/a$ could be as large as hundreds of thousands to millions of years \citep{Weir2007}, with the kinds of environmental fluctuations leading to resonant activation scaling accordingly. A major difference between the taxon-level and individual cases however is that $a$ could vary significantly within lineages, which is beyond the scope of the model presented here. 

The greater physical understanding of resonant activation developed here has facilitated simplified expressions for the extinction time to be derived from a heuristic perspective. Furthermore, a criterion for the existence of resonant activation is derived, indicating a dependence upon the chosen numerical parameters. Accordingly, we anticipate that the phenomenon may yet emerge within a broader range of dynamical systems than previously reported.

\section{Acknowledgements}
Much of this work was completed at the 2015 Geophysical Fluid Dynamics (GFD) Program at Woods Hole Oceanographic Institution. The GFD Program is supported by the US National Science Foundation (NSF) award OCE-1332750 and the Office of Naval Research. GRF was supported by NSF Award OCE-1155205. CRD's research was also supported in part by NSF Award DMS-1515161 and a Fellowship from the John Simon Guggenheim Foundation. CS thanks the NESSF16 Graduate fellowship for funding and thanks Woodward W. Fischer and Konstantin Batygin for enlightening discussions. We would furthermore like to thank the reviewers for providing well thought-out comments that significantly improved the paper.
%

%
%

\end{document}